\documentclass[aps,prb,twocolumn,floatfix,showpacs]{revtex4}
\usepackage{epsfig}

\begin{document}
\title{Numerical study of anharmonic vibrational decay in amorphous and paracrystalline silicon}

\author{Jaroslav Fabian}
\affiliation{Institute for Theoretical Physics, Karl-Franzens University, Universit\"{a}tsplatz 5, 8010 Graz, Austria}

\author{Joseph L. Feldman and C. Stephen  Hellberg}

\affiliation{Center for Computational Materials Science, Naval Research Laboratory, Washington DC 20375-5345}

\author{S. M. Nakhmanson}

\affiliation{Department of Physics, North Carolina State University, Raleigh, NC 27695-8202}

\begin{abstract}
The anharmonic decay rates of atomic vibrations in amorphous silicon (a-Si) and 
paracrystalline silicon (p-Si), containing small crystalline grains
embedded in a disordered matrix,  are calculated using realistic structural models.
The models are 1000-atom four-coordinated networks relaxed to a local 
minimum of the Stillinger-Weber interatomic potential. The vibrational decay rates 
are calculated numerically by perturbation theory, taking into account cubic anharmonicity as the perturbation.
The vibrational lifetimes for a-Si are found to be on picosecond time scales, 
in agreement with the previous perturbative and classical molecular 
dynamics calculations on a 216-atom model. The calculated decay rates for p-Si are
similar to those of a-Si. No modes in p-Si reside entirely on the crystalline
cluster, decoupled from the amorphous matrix. The localized modes with the largest (up to 59\%) 
weight on the cluster decay primarily to two diffusons. 
The numerical results are discussed in relation to a recent suggestion by
van der  Voort et al. [Phys. Rev. B {\bf 62}, 8072 (2000)] that long vibrational 
relaxation inferred experimentally may be due to possible crystalline nanostructures 
in some types of a-Si.  
\end{abstract}

\pacs{63.50.+x,65.60.+a}

\maketitle

\section{introduction}

The pioneering experiments by Dijkhuis and coworkers 
\cite{Scholten1993:PRB,Scholten1996:PRB,Voort2000:PRB,Voort2001:PRB}
explored transient dynamics of 
excited vibrational modes in a topologically disordered material--hydrogenated amorphous silicon. 
In these experiments nonequilibrium vibrational states were generated during relaxation
and recombination of optically excited electrons, and monitored with a probe laser (anti-Stokes
Raman spectroscopy) for transient behavior. The experimental results are surprising: 
Scholten et al. \cite{Scholten1993:PRB,Scholten1996:PRB} found that at low temperatures (2 K) and for
vibrational frequencies greater than 10 meV (maximum frequency in a-Si is about 70 meV) vibrations decay 
on time scales of tens of nanoseconds. Furthermore, the higher the vibrational frequency, the slower is the decay rate. 
In contrast, phonons in crystalline silicon decay on time scales of tens of picoseconds~\cite{Menendez1984:PRB}
 and the decay rates increase
with increasing frequency. The results of Scholten et al. were further confirmed 
by van der Voort et al., \cite{Voort2000:PRB} who suggested that the long lifetimes are
due to the microstructure of amorphous silicon. This suggestion was tested by van der Voort et al. 
\cite{Voort2001:PRB} by measuring the vibrational decay rates of a mixed amorphous-nanocrystalline
silicon, which was an amorphous hydrogenated silicon with a sizable fraction of nanocrystallites 
(with the diameter of $1-5$ nm). Even the mixed sample displayed nanosecond vibrational 
lifetimes, although the lifetimes appeared to decrease with increasing frequency.
A hypothesis was put forward~\cite{Voort2001:PRB} that the measured types of amorphous silicon
contain nanoscale regions with correlated (if not ordered) atoms, which, through enhanced
size quantization and localization of vibrational frequencies, inhibit anharmonic decay. 

These experimental results are at odds with the known theories of anharmonic vibrational decay
in disordered materials.~\cite{Fabian1996:PRL,Bickham1998:PRB,Bickham1999:PRB} 
In their so called ``fracton'' model, 
Alexander et al.~\cite{Alexander1983:PRB} assumed that the majority of vibrational
states in disordered systems are localized. This seemed to explain the above  experimental
findings since the anharmonic decay could be
drastically reduced by the extremely small likelihood of the overlap between three localized modes.
\cite{Jagannathan1989:PRB,Orbach1994:JPC} 
That the small probability of the overlap between three localized modes inhibits vibrational decay 
was disputed by Fabian and Allen~\cite{Fabian1996:PRL} who put forward a probabilistic 
scaling argument that the interaction between three localized modes would in fact be crucial 
for the anharmonic decay and cannot be neglected. Fabian later demonstrated~\cite{Fabian1997:PRB} 
the scaling argument on a one dimensional anharmonic chain with random spring constants,
and similar conclusions were reached recently by Leitner in a
study of heat flow in a one dimensional glass~\cite{Leitner2001:PRB} and vibrational energy transfer in helices of 
myoglobin.\cite{Leitner2001:PRL} Thus the fracton model, even if true in its premise of 
localization of the majority of the vibrational modes, does not explain the experiment.  
We note, however, that even the premise of the model is questionable, as it is in sharp
contrast to what is found in finite-size realistic models of glasses, which normally exhibit 
localization only in a small part of the spectrum. 

Numerical calculations of vibrational decay in glasses have been performed both by 
evaluating a perturbation formula~\cite{Fabian1996:PRL} and by classical 
molecular dynamics.\cite{Bickham1998:PRB,Bickham1999:PRB}
Perturbation theory was applied to the problem of anharmonic decay in glasses 
by Fabian and Allen \cite{Fabian1996:PRL} who computed the decay rates for a 216 atom model
of amorphous Si. The decay rates were found to be fractions of meV (that is, lifetimes are picoseconds),
in general increasing with increasing frequency. The anharmonic lifetimes of localized modes 
were similar to those of the extended modes, even in the case of a model
alloy Si$_{x}$Ge$_{1-x}$, where localized modes span more than a half of the
spectrum and the overlap between localized states becomes important.~\cite{Fabian1996:PRL}
Bickham and Feldman~\cite{Bickham1998:PRB} reported vibrational decay rates for 
selected modes of 216 and 4096 atom models of amorphous Si, using classical molecular 
dynamics. Their results agree with the perturbative calculation,
though the computed decay rates are somewhat greater due to the fact that molecular
dynamics takes into account all the anharmonic interaction, while the perturbative
calculation in Ref.~\onlinecite{Fabian1996:PRL} only cubic anharmonicity. 
In the calculation of Bickham and Feldman, a chosen vibrational mode was given a 
greater than average kinetic energy and was allowed to equilibrate while keeping 
the overall temperature constant. From the decay of the kinetic energy in time,
the mode decay rate was obtained. While the advantage of molecular dynamics over 
perturbation theory in calculating vibrational decay rates is that the full
anharmonic interaction is considered, the disadvantage is that the classical 
dynamics does not capture accurately the low temperature decay rates (for example, 
the rates computed by classical molecular
dynamics vanish at zero temperature,~\cite{Bickham1998:PRB} while in reality 
they are finite due to quantum effects~\cite{Fabian1996:PRL}).

The purpose of this paper is twofold: (i) to extend the previous numerical 
studies of perturbative anharmonic decay in homogeneous amorphous silicon (a-Si) to a larger system, and 
(ii) to calculate vibrational decay rates for a model of amorphous silicon --- paracrystalline silicon (p-Si) --- 
that includes nanocrystallites.
The larger system is a 1000-atom model of a-Si, prepared similarly to the previously used 
216-atom model.\cite{Fabian1996:PRL} The calculated decay rates display smaller statistical fluctuations 
and agree, on average, with those of the smaller model. 
Also, the larger system has a lower minimum frequency, allowing to 
calculate decay rates at the frequencies previously unattainable. 
Studying paracrystalline silicon, a material where small crystalline grains
are embedded in a disordered matrix, allows us to test the hypothesis of van der Voort 
\cite{Voort2001:PRB} regarding the structural origin of the anomalous long 
vibrational lifetimes in a mixed amorphous-nanocrystalline Si system. In our calculations we have 
used a 1000-atom (86 out of which belong to a single crystalline grain) model created
by Nakhmanson \textit{et.\ al.}~\cite{Nakhmanson2001:PRB} to simulate medium-range order in
amorphous silicon. We should point out, that, although providing a more realistic
subject for the verification of van der Voort's hypothesis than ``regular'' models for
a-Si, this simple model is neither an exact structural match to nanocrystalline Si sample
of Ref.~\onlinecite{Voort2001:PRB} (24\% crystalline fraction and 4.5 nm average grain diameter,
versus $\approx$ 10\% and 1 nm in the model) nor can it account for various other topological
defects present in real material. Still, if van der Voort's supposition were correct, we would observe 
inhibited decay rates of the modes which would be predominantly localized on the crystalline cluster in
the model. However, we do not find any modes localized exclusively on the cluster: one of the most 
cluster-localized modes has only 59\% weight on the cluster, and is therefore well coupled to the 
disordered matrix. It is  not surprising that  such modes have decay rates
similar to other localized states. 

We remark that the names amorphous and paracrystalline in reference to our models are a matter of 
terminological convenience rather than an attempt in classifying real materials. We refer
to a-Si as describing a homogeneous, continuous random network of silicon atoms, while p-Si
models are such networks filled with crystalline clusters. Real materials --- which are normally
termed amorphous silicon --- are likely of the p-Si type, containing nanoscale crystallites with
a distribution of sizes. \cite{Voyles2001:PRL}

In the following we first introduce the structural models of a-Si and p-Si and their
harmonic vibrational properties, then discuss the perturbative calculation of
anharmonic decay rates and present the results for the 1000-atom models of
a-Si and p-Si. Finally, we discuss our results with respect to the experiment.

\section{models}

\begin{figure}
\centerline{\psfig{file=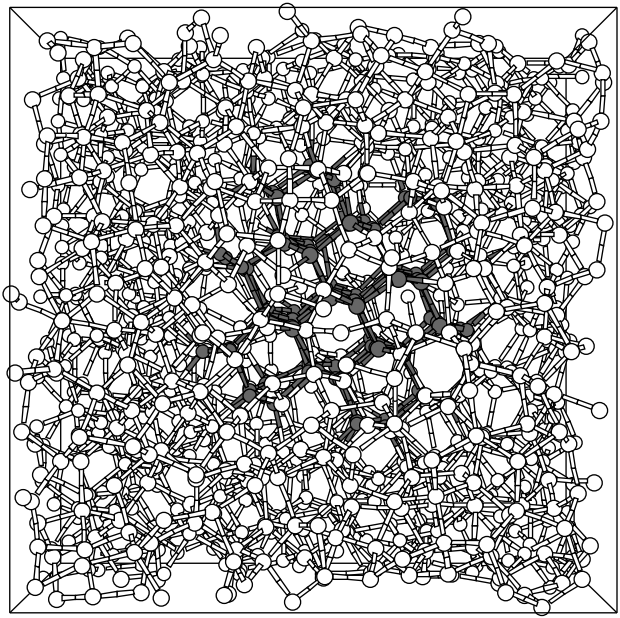,width=1\linewidth,angle=0}}
\caption{The 1000-atom model of p-Si. 
Atoms belonging to the crystalline grain are shown in grey; 
those in white form the amorphous matrix.
}
\label{fig:1}
\end{figure}

Both models employed in our studies were created with similar techniques:
the homogeneous model for a-Si was constructed using the WWW method~\cite{WWW1985:PRL} 
and the paracrystalline model with a variation~\cite{Nakhmanson2001:PRB} of the 
Barkema-Mousseau method.\cite{Barkema1996:PRL} (For a recent review
of modeling  continuous random networks see Ref.~\onlinecite{Mousseau2002:PMag}.)
The former model was also studied~\cite{Feldman1993:PRB}
for its harmonic properties within the framework of the Stillinger-Weber (SW)
potential\cite{Stillinger1985:PRB} prior to the present work.
Apart from the computational efficiency issues, the major difference between the methods of 
WWW and Barkema-Mousseau is the starting configuration used for the model construction: 
crystalline silicon (c-Si) is used in the former approach and a random close packed configuration in the latter.
Since neither of the starting models of this work employ the SW potential,
both models had to be relaxed with respect to SW prior to the decay times calculation.
This was done in an earlier work for the homogeneous model and in
the present work for the paracrystalline model through the use of a molecular dynamics quench,
where both temperature and 
virial pressure are set to zero in this stage of model preparation. The densities
for the two SW relaxed models are thus found to be slightly (three to four percent) less than that of
the density of c-Si, with the density of the paracrystalline model  being slightly (two percent) higher than
that of the homogeneous model. Changes in the atomic positions resulting from
the SW relaxation were found to be quite small and the view of the SW relaxed model of p-Si shown 
in Fig.~\ref{fig:1} is very close to the unrelaxed structure. In general
it is known that the SW potential produces relaxed structures that have
two to three percent five-fold coordinated atoms, even if the starting structures were
perfectly four-fold coordinated (which is the case for the paracrystalline model).
At variance with the optical properties of the models, this deviation from the perfect
four-fold coordination does not noticeably alter their vibrational properties and 
will not be discussed here in more detail. 

\begin{figure}
\centerline{\psfig{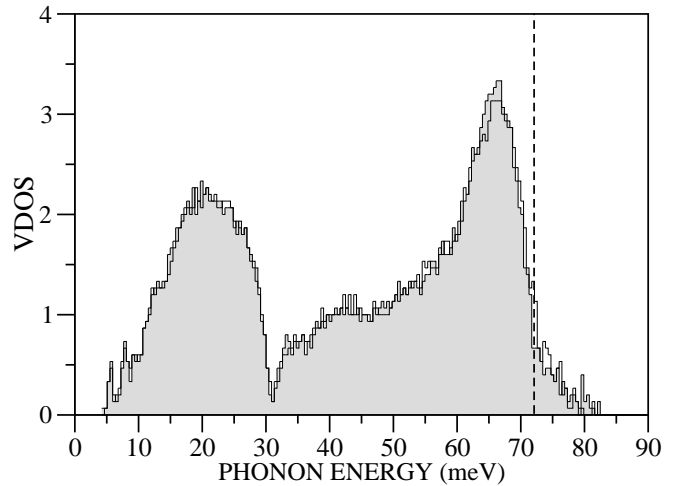}}
\caption{Vibrational density of states of the 1000-atom models of a-Si (shaded area) and p-Si
(line without shading).
Modes with $\omega > 72$ meV (indicated by the vertical line) are locons.}
\label{fig:2}
\end{figure}

\section{harmonic vibrations}

In the harmonic approximation  vibrational eigenfrequencies
$\omega(i)$  and eigenvectors ${\bf e}^{i}_a$ are computed
by diagonalizing the corresponding dynamical matrix (throughout the
paper symbols $j$, $k$, and $l$ will represent vibrational modes, while
$a$, $b$, and $c$ atoms). The results of numerical 
calculations from various groups~
\cite{Biswas1988:PRL,Schober1991:PRB,Schober1996:PRB,Taraskin2000a:PRB,Taraskin2000b:PRB,Marinov1997:PRB}
indicate that vibrational eigenstates in glasses
belong to one of four groups\cite{Fabian1996:PRL,Allen1999:PMB,Courtens2001:SSC}:
propagons, resonant modes, diffusons, and locons. Propagons are essentially
sound waves scattered by structural disorder. Resonances (sometimes called
quasilocalized modes) are modes temporarily trapped in topological
defects (in Si models these are the groups of undercoordinated atoms). Spatially,
a resonant mode has a large vibrational amplitude at the defect, but although
being rather weak, the mode does not decay exponentially (as a truly localized
mode would) outside of the defect region. In models of a-Si resonances often have 
frequencies below the lowest frequency (which is not zero due to the finite size) 
of the corresponding models of crystalline silicon. 
The low frequency spectral region of propagons and resonances is a subject
of great interest in relation to the so called boson peak,\cite{Nakayama2002:RPP} which was also
investigated in models of a-Si. \cite{Feldman1999:PRB,Nakhmanson2000:PRB}
Diffusons make up the majority of the spectrum. They have frequencies above 
the Ioffe-Regel limit (in principle, as suggested in
 Ref. \onlinecite{Feldman2002:JNCS}, there can be two such limits, one for transverse 
and one for longitudinal acoustic modes), so that they cannot
be characterized by a wave vector, but nevertheless are extended modes and contribute
to thermal conduction within Kubo theory as shown by the numerical calculations 
of Refs. \onlinecite{Allen1993:PRB,Feldman1993:PRB}.
Finally, locons are
localized modes in the sense of strong (Anderson) localization. In most 
studied models locons form only a small, high-frequency part of the spectrum. In our models
of a-Si many of the most localized locons reside in regions where there are five-fold coordinated
atoms. 

\begin{figure}
\centerline{\psfig{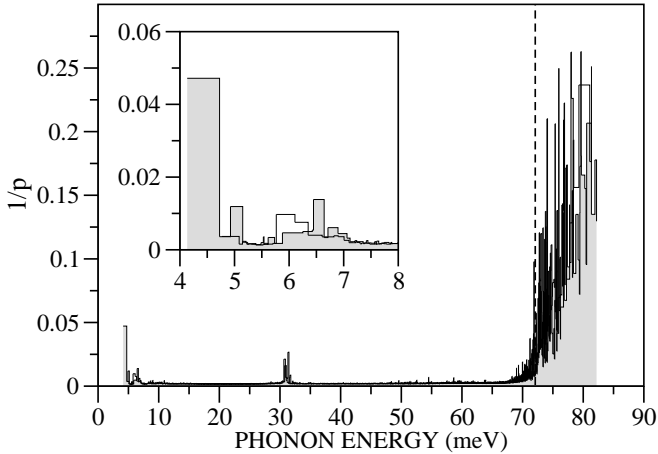}}
\caption{Inverse participation ratio $1/p$ of the vibrational states in the 1000-atom
models of a-Si (shaded area) and p-Si (line without shading). The modes with the frequencies above 72 meV (the vertical line)
can be considered localized. Quasilocalization occurs at low frequencies (resonant
modes) and around 30 meV which corresponds to the band edges. The inset is a detailed view
of the low-frequency region.}
\label{fig:3}
\end{figure}

Experimentally the character of the atomic vibrations in glasses has been 
studied by inelastic x-ray scattering in various glassy systems.
\cite{Foret1998:PRL,Ruocco1999:PRL,Pilla2000:PRL,Rat1999:PRL,Foret2002:PRB}
The general concensus seems to be that sound waves cease to
propagate early in the spectrum (for example, at 9 meV in densified
silica~\cite{Rat1999:PRL} and at 5 meV in amorphous selenium~\cite{Foret1998:PRL}).
This confirms the existence of the Ioffe-Regel crossover in glasses and the propagon picture.
The issue of the nature of the vibrational modes above the crossover 
frequency (where the models predict diffusons) has not been  resolved
experimentally yet. The recent experimental and theoretical progress is reviewed in
Ref.~\onlinecite{Courtens2001:SSC}.

In Fig. \ref{fig:2} we plot the calculated vibrational density of states (VDOS) for
the models of a-Si and p-Si. Both curves look very similar, which is in
agreement with the VDOS calculation of Ref.~\onlinecite{Nakhmanson2001:PRB} made
with a modified version of the SW potential.\cite{Vink2001:JNCS}
The calculated spectrum agrees rather well with the 
experimental one,~\cite{Feldman1993:PRB} except that the calculation
overestimates the highest
frequencies by about 15\%. This is a known artifact of the SW potential.

Localization properties of the modes can be judged from the participation
ratio $p(j)$, which indicates how many atoms ``participate'' in vibrational
eigenmodes $j$. Inverse participation ratio $1/p$ for a-Si and p-Si, as
a function of mode frequency is shown in Fig. \ref{fig:3}. The majority of vibrations in both 
models is delocalized,
with the localization transition taking place at around 72 meV (the mobility edge).
The modes around 30 meV and some modes below 10 meV appear to be localized too. The latter
are  resonant modes. The extended modes below about 15 meV
are propagons, while all the rest are diffusons (with possibly some longitudinal
propagons left~\cite{Feldman2002:JNCS} at small frequencies). The localization character in both a-Si and
p-Si models is similar. The presence of the crystalline cluster does not lead to additional
localized modes elsewhere in the spectrum.

\begin{figure}
\centerline{\psfig{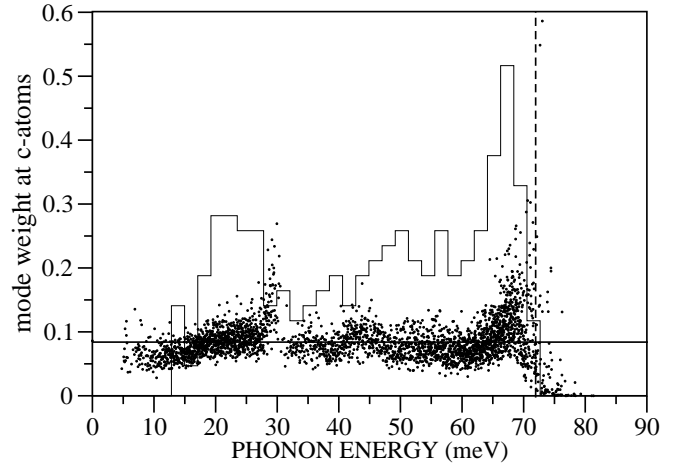}}
\caption{Weight of the modes at the crystalline cluster as a function of mode
frequency in the 1000-atom model of p-Si.
Plotted is the square of the atom displacement summed over
the  atoms forming the cluster. The horizontal line shows a weight of
0.086 ($8.6\%$) indicating an unbiased displacement pattern. The histogram
is VDOS for the cluster (see text).
}
\label{fig:4}
\end{figure}

In order to understand what fraction of each mode resides
on the crystalline cluster in the p-Si model, we compute the weight each mode
has on the cluster (that is, we sum $|{\bf e}^j_a|^2$ for each $j$ over all atoms 
$a$ from the cluster). The result is shown in Fig. \ref{fig:4}, together with
a histogram of VDOS of the cluster calculated by solving the dynamical equations
for the cluster atoms with the surrounding atoms held fixed. An unbiased mode 
has a weight of 0.086 (8.6\%), corresponding to the percentage of the atoms
making up the cluster. For all the modes below the mobility edge the weight
fluctuates around 0.086, showing no special affinity for the cluster. Localized
modes, as would be expected from their idiosyncratic character, can be localized
(fully or partially) on, through, or off the cluster. None of the modes is localized
fully on the cluster. There are 4 locons with the weight on the cluster 
of 30\% or greater, the maximum weight being that of 59\% for a mode with frequency
$\omega=$ 73.05 meV and participation ratio $p=13$. The second most localized mode
on the cluster has the frequency of 72.69 meV, the weight of 55\% and $p=12$. 
The third and fourth modes are more delocalized, having frequency (weight,$p$)
$\omega=70.67$ (31\%,160) and $\omega=71.12$ (30\%,117), respectively. 
All four modes lie in the mobility edge region. In addition to these, 
there are modes with frequencies around 30 meV which have enhanced affinity for the cluster
(see Fig. \ref{fig:4}). The weight of these modes at the crystalline cluster does not
exceed 30\%, but six modes have the weight between 20 and 30\%.

Harmonic vibrations in a-Si explain well many observed thermodynamic 
properties~\cite{Feldman2002:PMB} of the material, as well as kinetics
such as heat flow.\cite{Allen1993:PRB,Feldman1993:PRB} Anharmonicity 
does not directly affect heat flow in dielectric glasses, but is very
important in relaxing the perturbed vibrations to maintain local 
equilibrium (temperature gradient, to be specific).
More directly, anharmonicity affects thermal expansion and sound
attenuation. The 1000-atom model of a-Si was employed to demonstrate
the importance of thermal vibrations in both of these phenomena.
\cite{Fabian1997:PRL,Fabian1999:PRL} It was found that anharmonicity
is rather weak in a-Si, although somewhat stronger than in c-Si, primarily
due to strong anharmonicity of resonant modes. Indeed, resonant modes
show giant Gr\"{u}neisen parameters in the model, strongly enhancing the
effects of anharmonicity, although still within the limits of perturbation 
theory based on cubic anharmonicity.

\section{Vibrational lifetimes}

\begin{figure}
\centerline{\psfig{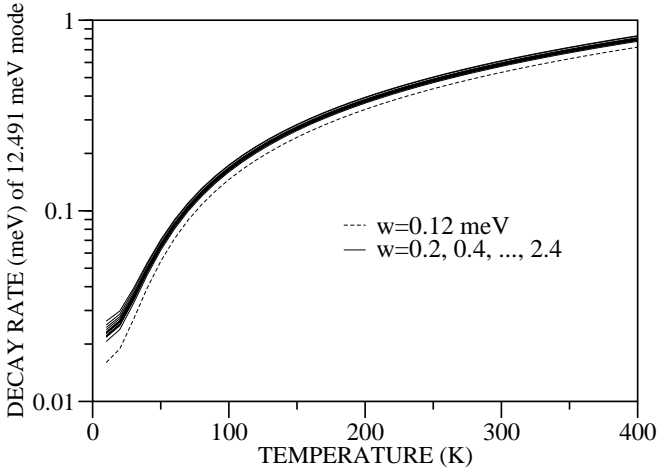}}
\caption{Calculated decay rate of the mode (a propagon) with frequency $\omega=12.49$ meV in a-Si as a function of
temperature for different widths $w$ of the rectangle function $\theta_w(\omega)$ representing the delta
function in Eq. \ref{eq:decay}. The curves are for $w$ equal 0.12 (dashed line) and 0.2, 0.4, ..., 2.4 meV (solid
lines), the order being not mirrored in the magnitude of the curves. The greatest decay rate is for $w=0.2$
meV, while the lowest for $w=1.4$ meV. The curve representing $w=1$ meV chosen in the calculation
is in the middle of the bunch.
}
\label{fig:5}
\end{figure}

Using cubic anharmonicity as the small perturbation to the harmonic Hamiltonian,
anharmonic decay rate $2\Gamma(j)$ of mode $j$ can be obtained from the 
formula\cite{Maradudin1962:PR,Cowley1963:AP} 
\begin{eqnarray} \nonumber
2\Gamma(j)&=& \frac{\hbar^2\pi}{4\omega(j)} \sum_{kl} \frac{|V(j,k,l)|^2}{\omega(k)\omega(l)} \\ 
  \nonumber & & \times  (\frac{1}{2} \left [1+n(k)+n(l)   \right ] 
             \delta\left [\omega(j)-\omega(k)-\omega(l)  \right ]   \\ \label{eq:decay} 
          & &+\left [ n(k)-n(l) \right ] \delta \left [\omega(j)+\omega(k)-\omega(l) \right ]
              ). 
\end{eqnarray}
Here $\omega(j)$ is the frequency of mode $j$, $n(j)$  is the mode occupation number 
given by $n(j)=\{\exp[\hbar\omega(j)/k_B T]-1\}^{-1}$ with $T$ denoting temperature,
and $V(j,k,l)$ is the matrix element of the cubic anharmonicity of the interatomic 
potential $V$ in the harmonic representation:  
\begin{equation} \label{eq:me}
V(j,k,l)=\sum_{abc}\sum_{\alpha \beta \gamma}
\frac{\partial^3 V}{\partial u_{a\alpha} \partial u_{b \beta} \partial u_{c\gamma}}
\frac{e^j_{a\alpha}}{\sqrt{m_a}} \frac{e^k_{b\beta}}{\sqrt{m_b}} \frac{e^l_{c\gamma}}{\sqrt{m_c}}.
\end{equation}
Greek symbols $\alpha$, $\beta$, and $\gamma$ stand for the cartesian coordinates of both the
atomic displacements ${\bf u}$ from the equilibrium positions, and of the normalized
vibrational eigenvectors ${\bf e}$. The atomic masses are denoted as $m$.
Anharmonic vibrational lifetimes are the inverse of the rates: 
\begin{equation}
\tau(j)=1/2\Gamma(j).
\end{equation}
In this paper we present decay rates in the units of meV. For conversion into lifetimes,
a decay rate of 1 meV is equivalent to a lifetime of  about 0.7 ps.

In Eq. \ref{eq:decay} the term with the temperature factor $1+n(k)+n(l)$ corresponds to 
the decay $j \rightarrow k+l$, while the term with $n(k)-n(l)$ represents the ``difference''
decay $j+k \rightarrow l$. Energy conservation is ensured by the delta functions. At low (down to
zero) temperatures the first term in Eq. \ref{eq:decay} dominates, giving rise to a constant $2\Gamma$, while both
terms are generally equally important at large temperatures, where $\Gamma \sim T$. In
crystals $V(j,k,l)$ vanishes unless the modes' momentum is conserved in the decay
process. In glasses, where lattice momentum itself is not a valid concept (except for
propagons and resonances), all the modes k and l from the spectrum contribute to $V(j,k,l)$
for a given $j$.

\begin{figure}
\centerline{\psfig{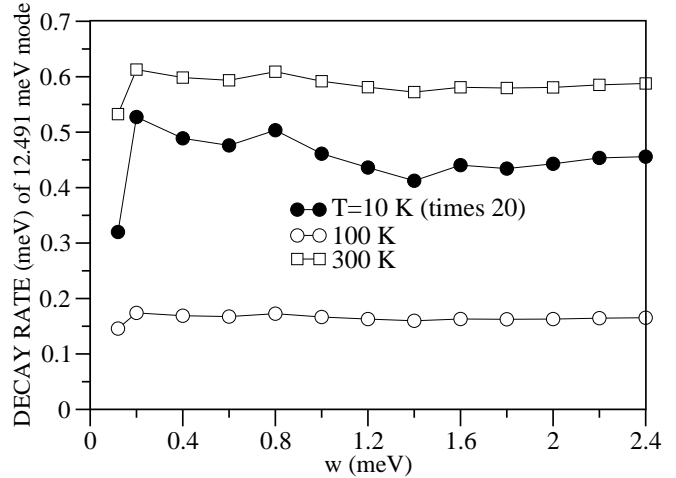}}
\caption{Calculated decay rate of the mode with frequency $\omega=12.49$ meV
in a-Si as a function of
$w$ at 10, 100, and 300 K.
}
\label{fig:6}
\end{figure}

Crucial for determining decay rates perturbatively from a finite-size model is the delta-function
regularization. We approximate $\delta (\omega) \approx \theta_{w}(\omega)$, where
$\theta_{w}(\omega)$ is a rectangle of width $w$ and height $1/w$ centered at $\omega=0$. In our calculations
with 1000 atoms we choose $w=1$ meV, which fits about 40 modes in the rectangle. 
The choice
of $w$ needs to be a compromise between good statistics and computer power. The statistics
is determined by both the number and ``similarity'' of the modes in a rectangle. If $w$ is
too large, the rectangle function will sample modes with distinct characteristics, not representing faithfully 
the modes of the chosen frequency. This problem is likely to be absent for diffusons, which do not differ
much on small spectral scales due to the absence of degeneracy (cf. Ref. \onlinecite{Fabian1997:PRL}), 
but may be relevant for propagons (which are mixed with
resonant modes) and locons (which are idiosyncratic). Fortunately, the averaging, first within
the rectangle and second, over the whole spectrum (see Eq. \ref{eq:decay}) makes the decay rates
quite insensitive to the choice of $w$, for a reasonable interval of values. In the earlier
calculation~\cite{Fabian1996:PRL} $w$ was chosen to be 0.4 meV for a 216-atom model, 
fitting about 4 modes in the rectangle. As we will see from the comparison of the two calculations 
in the following section, this choice
was already good enough, although it may have contributed somewhat to the statistical noise, 
especially at low temperatures and small frequencies. To illustrate the effect $w$ has on the decay rates, we
show in Fig. \ref{fig:5} the calculated rates of the mode with $\omega=12.49$ meV in a-Si, 
as a function of temperature, for selected $w$, ranging from $0.12$ meV (corresponding 
to about 4 modes per rectangle) to 2.4 meV (80 modes/rectangle). Except for the smallest
$w$, the results are grouped together with the dispersion of less than 10\% above 100 K. 
The greatest dispersion is at the lowest temperatures, where it reaches 25\%. (The low
temperature properties of the model do not describe well the real a-Si structure, because of the existence
of the minimum frequency of 4 meV in the model). Figure \ref{fig:6} shows the decay rate
for the same mode as a function of $w$, for selected temperatures. The rates become
reasonably insensitive to $w$ above 0.4 meV. The dispersion due 
to the sensitivity on $w$ is a factor contributing to the uncertainty
of the calculated values.

\section{amorphous silicon}

\begin{figure}
\centerline{\psfig{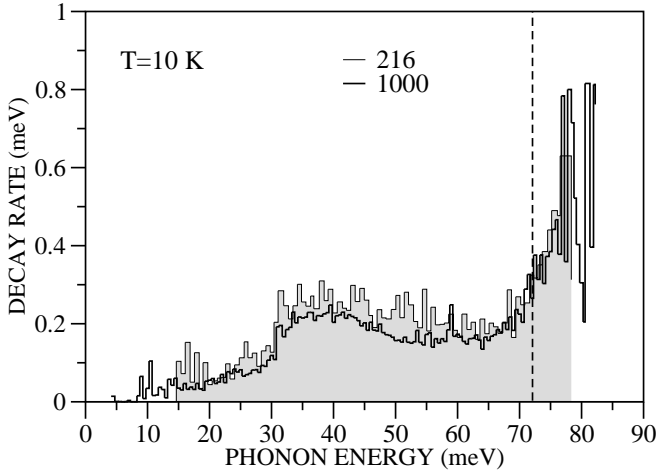}}
\caption{Calculated decay rates versus frequency for the 1000-atom model
at 10 K (thick line). For comparison the rates of the 216-atom model from Ref.
\onlinecite{Fabian1996:PRL} (not smoothed) are also shown (shaded area).}
\label{fig:7}
\end{figure}

\begin{figure}
\centerline{\psfig{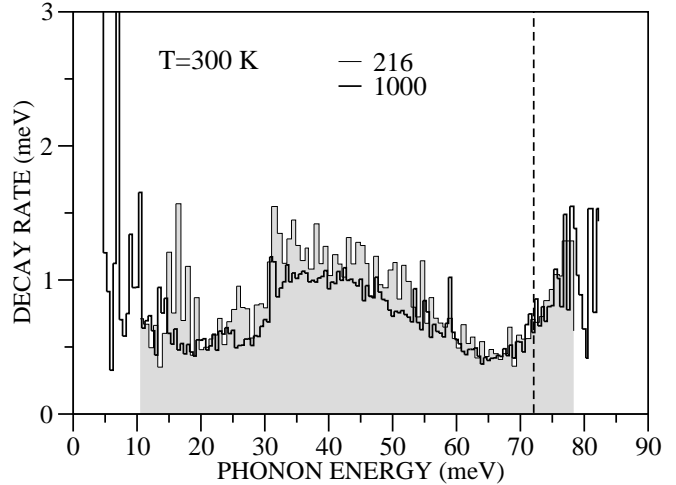}}
\caption{Calculated decay rates versus frequency for the 1000-atom model
at 300 K. For comparison the rates of the 216-atom model from Ref.
\onlinecite{Fabian1996:PRL} (not smoothed) are also shown (shaded area).}
\label{fig:8}
\end{figure}

We now present the calculated decay rates for the 1000-atom model of a-Si. 
Due to the computational power limitations we have sampled the spectrum
uniformly with about 200 modes for which we computed $2\Gamma$. 
The calculated decay rates are presented as a function of the
mode frequency for two different temperatures: 10 K in Fig. \ref{fig:7} and 
300 K in Fig. \ref{fig:8}. For comparison the previous calculations~\cite{Fabian1996:PRL} 
on a 216-atom model of a-Si are included. Overall, the decay rates for 
the two models agree. The rates are on the order of meV (picosecond lifetimes). 
Perturbation theory is thus valid for all the sampled modes with the exception of 
few in the lowest part of the spectrum at 300 K (see below).
As was shown in Ref. \onlinecite{Fabian1996:PRL} the decay rates as a function of 
frequency at 10 K (and at low temperatures, generally) follow the joint density
of states [$\sum_{kl}\delta(\omega(j)-\omega(k)-\omega(l))$] which counts, 
for a chosen mode $j$, the number of combination decay 
possibilities $j\rightarrow k+l$ with the constraint of energy conservation. 
At larger temperatures one must add the number of difference decay channels
$j\rightarrow k-l$ to reproduce, qualitatively, the calculated $2\Gamma(\omega)$.
More detailed physics of the anharmonic decay in glasses and especially 
the statistics of the decay matrix elements can be found in Ref. 
\onlinecite{Fabian1996:PRL}.

There are several features which make the calculated decay rates for the 216-atom 
and 1000-atom models somewhat different. The first is the overall 
reduction in noise for the 1000-atom model (the data are not
smoothed as was done in Ref. \onlinecite{Fabian1996:PRL}). The reason is
both the greater model size (spectral averaging) and greater $w$ (rectangle
averaging). Note that the observed noise in the spectrally resolved $2\Gamma$
is about 10\% or less, consistent with a dispersion of the decay
rates with $w$, discussed in the previous section.  
Second, the calculated rates for the 1000-atom model are somewhat
smaller than those of the 216-atom model, that is, the latter model appears
to be slightly more anharmonic. This is at variance with the calculation of
thermal expansion~\cite{Fabian1997:PRL} where the 216-atom model seems less
anharmonic. The latter difference probably can be explained by the anomalously
large negative mode Gruneisen parameters of the low frequency
resonance modes of the 1000 atom model, as the thermodynamic Gruneisen parameter
depends on an average mode Gruneisen parameter at high temperatures. We note that
the structural models differ in other ways: the smaller model is
more topologically constrained,\cite{Fabian1997:PRL} has smaller
energy/atom, and has higher density than the 1000-atom model.
Third, the calculated rates of the 1000-atom model extend to a lower frequency region as the
minimum frequency of the model is smaller than that of the 216-atom model. 
Finally, some low-frequency modes (resonances) at 300 K exhibit giant decay rates, comparable
to the modes' frequencies. These rates are in fact invalid, since they 
are not consistent with perturbation theory. However, they indicate what
may be expected from a full anharmonic calculation (for example by 
molecular dynamics). This important physics issue will be discussed elsewhere.

\begin{figure}
\centerline{\psfig{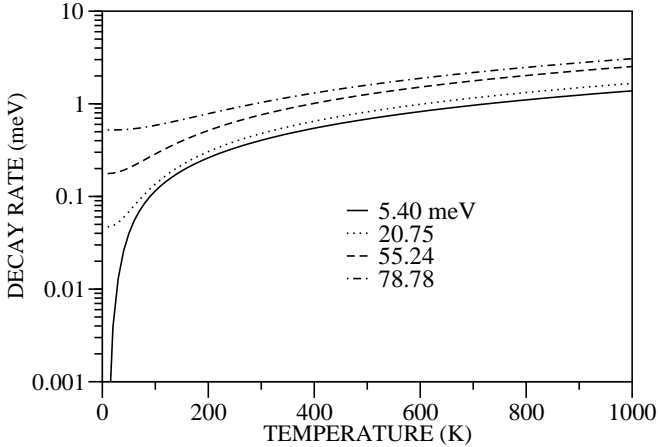}}
\caption{Calculated decay rates of selected modes in a-Si versus temperature. 
The lines are labeled according to modes' frequencies in meV. The lowest frequency mode
is a propagon, the following two are diffusons (acoustic-like and optic--like)
and the highest frequency mode is a locon.
}
\label{fig:9}
\end{figure}

In Fig. \ref{fig:9} we plot the temperature dependence of the decay rates of
selected modes. We show the temperature dependence  for
a propagon, an acoustic-like and an optic-like diffuson, and a locon. The low-frequency
propagon has a divergent lifetime (decay rate vanishes) as temperature
decreases to zero, since there are no two modes into which it could decay, due to the
energy conservation constraint and the existence of the minimum-frequency mode. 
All the other modes have constant decay rates at small temperatures.
The constant goes smoothly to a linear function at large temperatures,
which is due to the fact that the population density of thermal phonons increases
linearly with temperature.

\section{paracrystalline silicon}

\begin{figure}
\centerline{\psfig{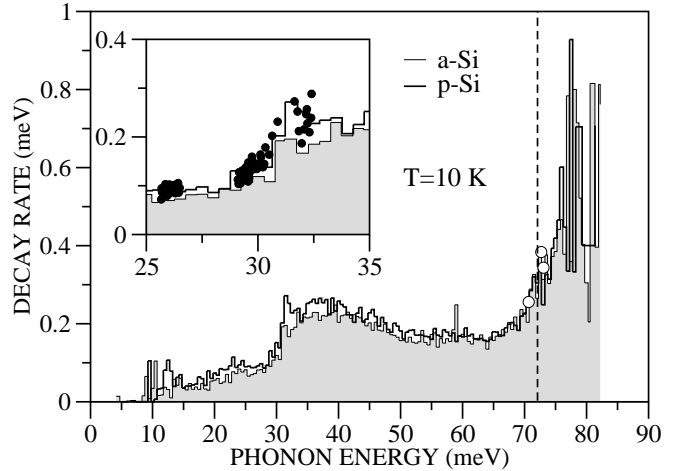}}
\caption{Calculated decay rates of the 1000-atom model of p-Si at 10 K.
For comparison, the rates for the 1000-atom model of a-Si are also shown (shaded area).
The empty circles are for three modes in the mobility region with more than 30\% weight on the
crystalline cluster, while the inset shows the decay rates (filled circles) of modes around 30 meV, which
have large affinity (weight up to 30 \%) for the cluster.
}
\label{fig:10}
\end{figure}

\begin{figure}
\centerline{\psfig{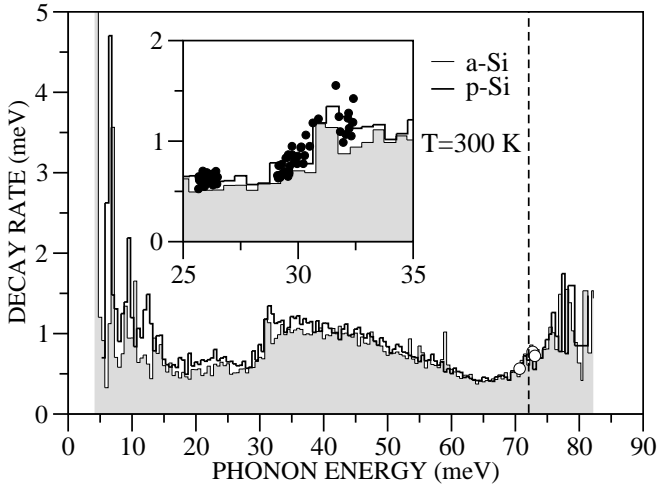}}
\caption{Calculated decay rates of the 1000-atom model of p-Si at 300 K.
For comparison, the rates for the 1000-atom model of a-Si are also shown (shaded area).
The empty circles are for three modes with more than 30\% weight on the
crystalline cluster, and the inset plots the decay rates (filled circles) of 
the modes around 30 meV with large weight on the cluster.}
\label{fig:11}
\end{figure}

The results for the 1000-atom model of p-Si are shown in Figs. \ref{fig:10}
and \ref{fig:11}, which plot $2\Gamma$ as a function of mode frequency. 
For comparison we also present the data for a-Si discussed in the previous 
section.  The results are quantitatively similar for both models.
There are no anomalous decay rates appearing in the spectrum of p-Si
which would be due to the crystalline cluster. In addition to 
the sampling modes, we have computed the decay rates
specifically for three modes in the mobility edge region with the weight at the 
crystalline cluster greater
than 30\%. They are presented in Figs. \ref{fig:10} and \ref{fig:11} by
empty circles. The decay rates of these modes have the same magnitude as
those of the other locons. Finally, in the insets of the two figures we show 
the decay rates of modes with frequencies around 30 meV, the region of
especially high affinity for the cluster (see Fig. \ref{fig:4}). Decay
rates of more than 80 modes in that spectral region are plotted. Although
many of the modes have large weight (some of them up to 30\%) on the
cluster, most are unbiased. The fact that $2\Gamma$ of all of these modes are
similar in magnitude at different temperatures implies no special decay
behavior for the modes of strong affinity for the cluster.

\begin{figure}
\centerline{\psfig{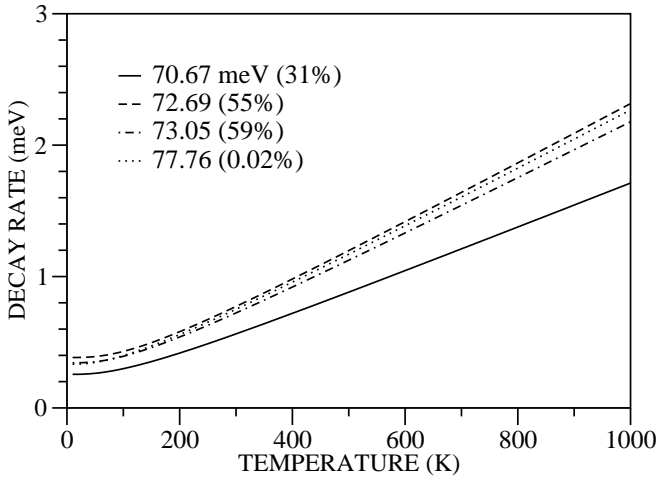}}
\caption{Calculated decay rates for selected locons in p-Si. The curves are
labeled according to frequency in meV. The numbers in the brackets show the
modes' weight on the crystalline cluster.
}
\label{fig:12}
\end{figure}

\begin{figure}
\centerline{\psfig{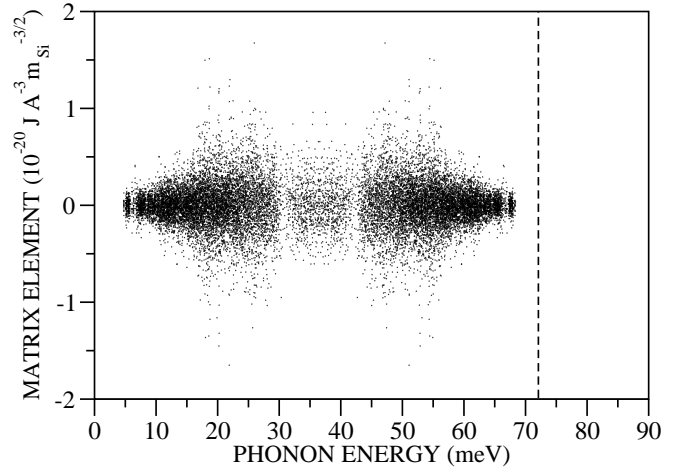}}
\caption{Calculated matrix element V(j,k,l) of p-Si for mode $j$ maximally localized (59\%) on the
crystalline cluster [$\omega(j)=73.05$ meV] as a function of $\omega(k)$. Shown are data for
$k$ and $l$ obeying energy conservation: $|\omega(j)-\omega(k)-\omega(l)| < w/2$, where $w=0.2$ meV
(taken to be smaller than $w=1$ meV used in the calculation, to get a manageable graphics 
size). 
}
\label{fig:13}
\end{figure}

Figure \ref{fig:12} shows the temperature dependence of three modes with 
more than 30\% weight on the crystalline cluster. The modes have frequencies
(weight,p) 70,67 meV (31\%,160), 72.68 meV (55\%,12), and 73.05 meV (59\%,13).
In addition, the figure plots the decay rate of a ``normal'' locon with
$\omega=77.76$ meV (0.02\%,8), residing outside of the cluster. No 
unusual features are observed in the temperature dependence of the
decay rates of these modes. Finally, in Fig. \ref{fig:13} we plot the anharmonic matrix elements $V(j,k,l)$
of the combination decay $j \rightarrow k+l$ for the maximally localized mode on 
the cluster, with frequency 73.05 meV to visualize the decay channels. The figure shows that 
the dominant channel is a decay into two diffusons. Decay into a propagon and a diffuson
(the points in Fig. \ref{fig:13} below 15 meV and above 58 meV) is somewhat 
less important. The corresponding matrix elements are much smaller. This may
be related to the fact that propagon's weight on the cluster is systematically
lower than 8.5\% (see Fig. \ref{fig:4}). The diffusons' weight on the cluster 
is much more more scattered, with a significant number of diffusons having
the weight of 8.5\% and more. Decay into another locon and a propagon is forbidden by energy conservation.

\section{conclusions}

We have calculated anharmonic decay rates of 1000-atom models of a-Si and p-Si 
using perturbation theory with cubic anharmonicity in the interatomic potential.
The results for a-Si are in agreement with the previous perturbative calculations on a smaller
model, as well as with a molecular dynamics calculation.  The results reiterate
the previous findings that the vibrational lifetimes are on the picosecond time
scales, generally increasing with increasing frequency. The decay rates of locons
are idiosyncratic, but are by no means inhibited. Calculated decay rates of
p-Si are similar to those of a-Si, showing little sensitivity to structural properties.
These findings disagree with the interpretation of recent experiments which find
decay rates on the order of nanoseconds and somewhat greater sensitivity to structural
properties. 

The explanation that we offer to account for these discrepancies is that the calculation 
and experiment refer to two different things. First, as we have pointed out earlier,
simple (and at such scale usually over-relaxed) models like a continuous random network 
type WWW model or a similar model containing a crystalline grain used in this study can not 
faithfully reproduce a broad range of various topological features --- some or combinations 
of which may be responsible for increased decay times observed in the experiment --- present 
in a real material. Second, in our calculations only ``perturbative'' decay rates,
where a small (infinitesimal) population of a single mode goes out of
equilibrium are computed. The experiments measure the decay of vibrational states excited over a large
portion of the spectrum. Furthermore, the laser excitation produces phonon populations too far
off the equilibrium to be called small perturbations. In Ref.~\onlinecite{Voort2000:PRB},
for example, the excited phonon population $n$ lies between 0.03 and 0.3, which, for a
mode with frequency, say, 50 meV corresponds to an effective temperature of 160 and
400 K, respectively. This is huge compared to 2 K at which the samples are held.
A numerical investigation of Bickham~\cite{Bickham1999:PRB} indeed shows that a strong
perturbation of the vibrational spectrum of a-Si can relax on a 100 ps time scale,
compared to 10 ps for a weak perturbation.  In addition to pure vibrational relaxation,
it is also likely, as suggested by Bickham and Feldman,\cite{Bickham1998:PRB}
that correspondingly large local deviations in the atomic displacements cause local structural 
rearrangements which may relax to local metastable minima while emitting phonons.
It is possible that these rearrangements occur on nanosecond timescales.

We conclude that current simple models for a-Si in combination with presented
above methods of analysis do not provide an answer to the question why vibrations in real
material under the specific experimental conditions appear to decay over such long time scales. 
A better answer may be obtained by studying larger and more realistic models when computational 
power required to make such calculations tractable becomes available. However, considering
the consistent results obtained thus far with different models (of sizes from 216 to 4096 atoms) and
different techniques (perturbation theory and molecular dynamics) it is rather unlikely
that the interpretation of the experimental findings in terms of the (pure) vibrational
lifetimes (that is, without considering structural relaxation and possibly photoexcited-electron 
recombination processes) will be validated by investigating larger models.

\acknowledgments{We thank Phil Allen for useful discussions and M.
van der Voort for suggesting this calculation.}

\bibliographystyle{apsrev}
\bibliography{glass}

\end{document}